\documentclass[iop,apj]{emulateapj}
\usepackage{amsmath,amssymb,amstext}

\submitted{Received 2016 October 14; revised 2016 December 19; accepted 2016 December 20}
\journalinfo{The Astrophysical Journal}

\usepackage[breaklinks,colorlinks,citecolor=black,linkcolor=black]{hyperref} 

\usepackage[all]{hypcap} 

\usepackage{aas_macros}
\usepackage{natbib}
\newcommand{\Msun}{\ensuremath{M_\odot}}

\newcommand{\Ni}{\ensuremath{^{56}\mathrm{Ni}}}
\newcommand{\Co}{\ensuremath{^{56}\mathrm{Co}}}

\newcommand{\Mej}{\ensuremath{M_\mathrm{ej}}}
\newcommand{\Eej}{\ensuremath{E_\mathrm{ej}}}

\shorttitle{Magnetars mimicking \Ni}
\shortauthors{Moriya, Chen, \& Langer}

\begin{document}

\title{Properties of magnetars mimicking \Ni-powered light curves \\ in Type~Ic superluminous supernovae}
\author{Takashi J. Moriya$^{1}$, Ting-Wan Chen$^{2}$, and Norbert Langer$^3$}
\affil{$^1$Division of Theoretical Astronomy, National Astronomical Observatory of Japan, National Institutes for Natural Sciences, \\ 2-21-1 Osawa, Mitaka, Tokyo 181-8588, Japan; takashi.moriya@nao.ac.jp}
\affil{$^2$Max-Planck-Institut f{\"u}r Extraterrestrische Physik, Giessenbachstra\ss e 1, D-85748 Garching, Germany}
\affil{$^3$Argelander Institute for Astronomy, University of Bonn, Auf dem H\"ugel 71, 53121 Bonn, Germany}

\begin{abstract}
Many Type~Ic superluminous supernovae have light-curve decline rates after their luminosity peak
which are close to the nuclear decay rate of \Co, consistent with the interpretation that they are
powered by \Ni\ and possibly pair-instability supernovae.
However, their rise times are typically shorter than those expected
from pair-instability supernovae, and Type~Ic superluminous supernovae
are often suggested to be powered by magnetar spin-down.
If magnetar spin-down is actually a major mechanism to power
Type~Ic superluminous supernovae, it should be able to produce 
 decline rates similar to the \Co\ decay rate rather easily.
In this study, we investigate the conditions for magnetars under which their spin-down energy input
can behave like the \Ni\ nuclear decay energy input.
We find that an initial magnetic field strength 
within a certain range is sufficient to keep the magnetar energy deposition
within a factor of a few of the \Co\ decay energy for several hundreds of days.
Magnetar spin-down needs to be by almost pure dipole radiation
with the braking index close to 3 to mimic \Ni\ in a wide parameter range.
Not only late-phase \Co-decay-like light curves, but also
rise time and peak luminosity of most \Ni-powered light curves can
be reproduced by magnetars.
Bolometric light curves for more than 700~days are required
to distinguish the two energy sources solely by them.
We expect that more slowly-declining superluminous supernovae with short rise times
should be found if they are mainly powered by magnetar spin-down.
\end{abstract}

\maketitle

\section{Introduction}\label{sec:introduction}
Supernovae (SNe) were first thought to be the birth place of neutron stars by \citet{baade1934}.
Currently, most neutron stars are believed to be born
during core collapse of massive stars leading to the core-collapse SNe.
If newly-born neutron stars rotate rapidly and have strong magnetic
fields (so-called ``magnetars''), they can affect subsequent
observational properties of exploding stars.
This is because the huge rotational energy of neutron stars
can be radiated on a short timescale if they are strongly
magnetized \citep[e.g.,][]{ostriker1971magsn}.
Magnetars are suggested to be engines of
several energetic events observed in the Universe, such as
gamma-ray bursts (GRBs) and SNe with large explosion energies
\citep[e.g.,][]{usov1992maggrb,thompson2004maghngrb,maeda2007mag05bf,mazzali2006nsdriven06aj,mazzali2014grbsnmag,komissarov2007mag,burrows2007mag,greiner2015ulgrbslsn,metzger2015magtran,moriya2016metzger}.

In particular, magnetar spin-down has been suggested to be a candidate energy
source to power superluminous SNe (SLSNe)
\citep[e.g.,][]{kasen2010mag,woosley2010mag,dessart2012magslsn,chatzopoulos2013anachi,inserra2013slsntail,nicholl2013ptf12damnature,nicholl2015slsndiv,mccrum2014slsnps111ap,metzger2014ionbreak,metzger2015magtran,chen2015ptf12damhost,moesta2015mag,kasen2016magbreakout,wang2016triple13ehe,bersten2016asassn15lh,sukhbold2016woosley}.
SLSNe are a recently-recognized
class of SNe whose peak luminosity is more than $\sim 10$ times higher than
that of canonical SNe (see \citealt{gal-yam2012slsnrev} for a review).
Among several spectral types of SLSNe, we focus here on Type~Ic SLSNe
whose optical spectra do not generally show hydrogen features
(e.g., \citealt{quimby2011slsn}, see also a recent discovery by \citealt{yan2015iptf13ehe} and its interpretation by \citealt{moriya2015slsnhstrip}).

One of the first discovered Type~Ic SLSNe, SN~2007bi, is suggested
to originate from a pair-instability SN (PISN) powered by a huge amount of \Ni\
which is synthesized when they explode (\citealt{gal-yam2009sn2007bi}, but see also \citealt{young2010sn2007bi,moriya2010sn2007bi}).
This is because its light-curve (LC) decline rate after the LC peak
is consistent with the nuclear decay rate of \Co\
and strong nebular Fe emission lines are also observed, as expected in PISNe
\citep[e.g.,][]{kasen2011pisn,dessart2013pisn,kozyreva2014pisnlc,whalen2014pisn,chatzopoulos2015rotpisn}.

However, subsequent theoretical and observational studies suggest that
the short LC rise times and little line blanketing of slowly-declining Type~Ic SLSNe
like SN~2007bi are inconsistent with \Ni-powered PISNe and
prefer the magnetar-powered model
\citep[e.g.,][]{dessart2012magslsn,inserra2013slsntail,nicholl2013ptf12damnature,nicholl2015slsndiv,nicholl2016sn2015bn,mccrum2014slsnps111ap,chen2015ptf12damhost,jerkstrand2016pisnnebular,jerkstrand2016slsnnebular,mazzali2016slsn}
or other power sources 
\citep[e.g.,][]{chevalier2011irwin,ginzburg2012csmsn,moriya2012dip,ouyed2012sn2006gyquarknova,chatzopoulos2013anachi,dexter2013kasen,baklanov2015ptf12dam,sorokina2016slsnicsm,nicholl2015lsq14bdqclearbump,wang2016triple13ehe}.
In addition, there also exist many Type~Ic SLSNe with fast LC declines
that cannot be explained by the \Ni\ power
\citep[e.g.,][]{pastorello2010sn2010gx,chomiuk2011panslsn,inserra2013slsntail,howell2013slsn,nicholl2015slsndiv}.
The magnetar model can provide a unified explanation for 
fast-declining and slow-declining Type~Ic SLSNe \citep[e.g.,][]{kasen2010mag}.

Several environmental studies of SLSNe have been carried out, 
and low metallicity and high specific star-formation rate are commonly found in SLSN host galaxies \citep[e.g.,][]{neill2011slsnhost,chen2013sn2010gxhost,chen2015ptf12damhost,lunnan2014slsnhost,thoene2015ptf12damhost,leloudas2015slsnhost,angus2016slsnhost,perley2016slsnhost,japelj2016slsngrbhosts}. 
However, they are
still insufficient to distinguish different possible powering mechanisms.
\citet{chen2016toogoodrelation} recently find a possible relation between the initial spin obtained by
fitting SLSN LCs assuming they are powered by magnetars and their host metallicity.
If confirmed, this relation may prefer the magnetar model, but further studies are required.

One interesting feature of slowly-declining Type~Ic SLSNe is that their LC decline rates 
are often consistent with the \Co\ decay rate for a long time
\citep[e.g.,][]{nicholl2016sn2015bn,chen2015ptf12damhost,papadopoulos2015decamslsn,yan2015iptf13ehe,lunnan2016ps114bjlonghpoorslsn,vreeswijk2016slsnearlyexcess}.
If we just examine the late-phase LCs, the simplest way to explain them is the \Co\ decay.
If magnetars are actually powering most of Type~Ic SLSNe including
slowly declining ones,
it would be relatively easy for them to behave like the decaying \Co\ for a long time.
In this paper, we investigate how well magnetars can behave like \Ni\ in
powering SNe. We examine whether magnetars can actually behave like \Ni\
or not, and derive conditions for magnetars to mimic the \Ni\ nuclear decay.

The rest of this paper is organized as follows.
At first in Section~\ref{sec:mimi}, we investigate conditions
under which magnetar spin-down energy can be similar to \Co\
decay energy during late phases. Then, we also look into early LC
properties (rise time and peak luminosity) in Section~\ref{sec:early}
to see if magnetar-powered SNe can be similar to \Ni-powered SNe
in both early and late phases.
We discuss our results in Section~\ref{sec:discussion} and
summarize our conclusions in Section~\ref{sec:conclusions}.

\section{Mimicking conditions at late phases}\label{sec:mimi}
We first investigate conditions under which magnetars can have a similar energy
input to the \Co\ decay at late phases
($t\gtrsim 100$~days, where $t$ is the time since the explosion).
\Co\ in SN ejecta appears as a result of the decay of \Ni\
synthesized at the explosion. At $t\gtrsim100$ days, 
the energy input from the \Ni\ decay is negligible
because of its short decay time (8.7~days). Thus, we approximate
the energy input from the nuclear decay as
\begin{equation}
 L_\mathrm{\Co}=1.5\times 10^{43}M_\mathrm{\Ni 1}\exp\left(-\frac{t}{111\ \mathrm{days}}\right)~\mathrm{erg~s^{-1}},
\label{copowerlate}
\end{equation}
where $M_\mathrm{\Ni 1}$ is the initial \Ni\ mass $(M_\mathrm{\Ni})$
in units of $\Msun$.

The total rotational energy $E_p$ available in a neutron star to power SN LCs is
\begin{equation}
E_p=\frac{1}{2}I_\mathrm{NS}\Omega_i^2
\simeq 2\times 10^{52}P_\mathrm{ms}^{-2}~\mathrm{erg},
\end{equation}
where $I_\mathrm{NS}\simeq 10^{45}~\mathrm{g~cm^2}$ is the momentum of inertia
of a neutron star, $\Omega_i$ is its initial angular velocity, and
$P_\mathrm{ms}$ is its initial rotational period $(2\pi/\Omega_i)$
scaled with 1~ms.
We assume that the rotational energy
is lost by dipole radiation during the spin-down timescale of
\begin{equation}
t_p=\frac{6I_\mathrm{NS}c^3}{B^2R_\mathrm{NS}^6\Omega_i^2}
\simeq 4.1\times 10^{5}B_{14}^{-2}P_\mathrm{ms}^{2}~\mathrm{sec},
\label{eqtp}
\end{equation}
where $R_\mathrm{NS}\simeq 10$~km is the neutron star radius,
$c$ is the speed of light, $B$ is the initial neutron star magnetic
field strength, and $B_{14}$ is $B$ scaled with $10^{14}$~G
\citep[e.g.,][]{gunn1969mag,contopoulos1999magspindown,kasen2010mag}.
In Eq.~(\ref{eqtp}), we
assume that the angle between the magnetic dipole and the rotational axis
is $45^{\circ}$.
The energy deposited from magnetar spin-down is approximated as
\begin{equation}
L_\mathrm{mag}=\frac{\left(l-1\right)E_p}{t_p}
\left(1+\frac{t}{t_p}\right)^{-l}.
\end{equation}
The temporal index $l$ is 2 if magnetar
spin-down is purely through dipole radiation, i.e., a braking index of 3.

We here investigate conditions under which the energy supply from magnetar 
spin-down remains within a factor of $a$ of the \Co\ decay energy
at late phases, i.e.,
\begin{equation}
 a^{-1}<\frac{L_\mathrm{mag}}{L_\mathrm{\Co}} < a.
\label{eqmimic}
\end{equation}
We first consider the case of $l=2$.
In late phases of SNe as we are interested in here,
$t\gg t_p$ is satisfied 
(see Section \ref{sec:early} for discussion on early phases).
Thus, we can approximate the magnetic spin-down energy as
\begin{equation}
 L_\mathrm{mag}
\rightarrow E_pt_pt^{-2}=8.2\times 10^{57} B_{14}^{-2} t^{-2}\
\mathrm{erg~s^{-1}}.
\label{magpowerlate}
\end{equation}
Using the luminosity inputs at late phases
(Eqs. \ref{copowerlate} and \ref{magpowerlate}),
we obtain the following condition from Eq.~(\ref{eqmimic}) in the cgs unit:
\begin{equation}
 5.4\times 10^{14}a^{-1}t^{-2}e^{\frac{t}{111\ \mathrm{days}}}
< M_\mathrm{\Ni 1}B^2_{14} <
 5.4\times 10^{14}at^{-2}e^{\frac{t}{111\ \mathrm{days}}}.
\label{mimiccond}
\end{equation}
Equation (\ref{mimiccond}) indicates that only the initial magnetic field strength needs to be in a certain range for magnetars
to mimic the \Co\ decay when $l=2$.

\begin{figure}
 \begin{center}
  \includegraphics[width=\columnwidth]{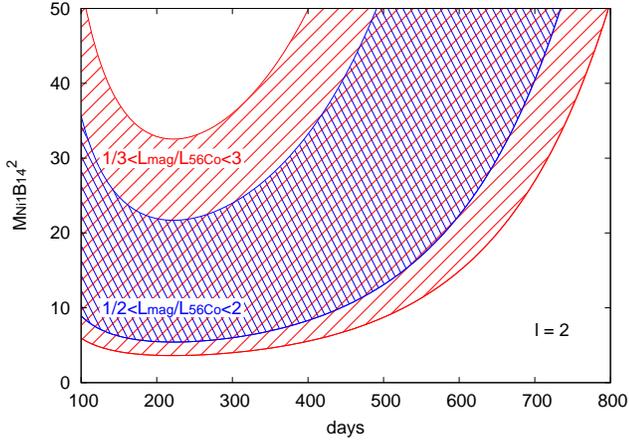} 
 \end{center}
\caption{
Ranges of $M_\mathrm{\Ni 1}B_{14}^2$ where pure dipole magnetar
spin-down energy ($l=2$, a braking index of 3)
is within factors of 2 or 3 of the \Co\ decay energy.
}\label{fig:mb2}
\end{figure}

\begin{figure}
 \begin{center}
  \includegraphics[width=\columnwidth]{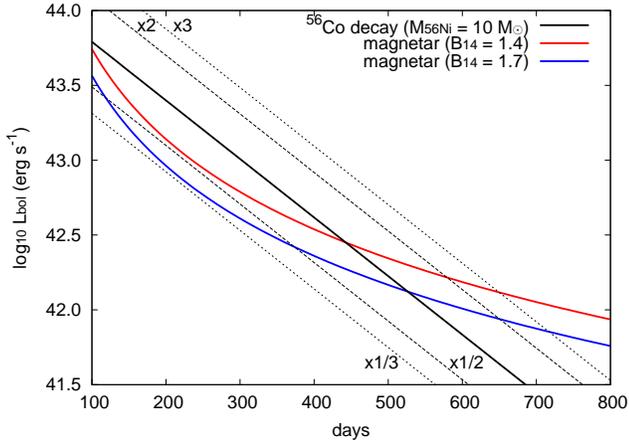} 
 \end{center}
\caption{
Comparison between magnetar spin-down energy and \Co\ decay
 energy.
}\label{fig:comp}
\end{figure}

\begin{figure}
 \begin{center}
  \includegraphics[width=\columnwidth]{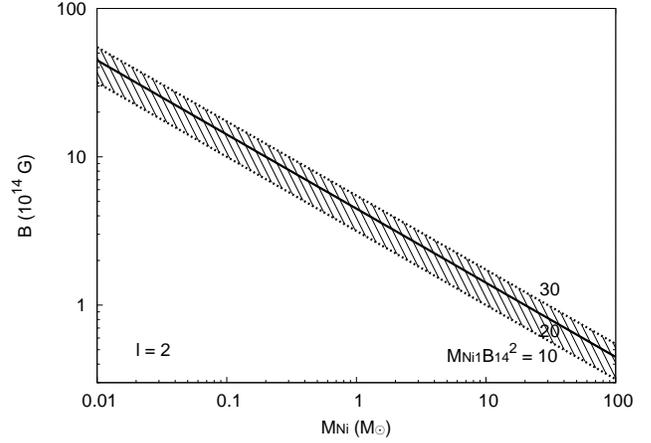} 
 \end{center}
\caption{
The parameter region where $10\lesssim M_\mathrm{\Ni 1}B_\mathrm{14}^2\lesssim 30$ is satisfied. The late-phase LCs from a given amount of \Ni\ can be mimicked by magnetars with $B$ in the shaded region if a braking index is 3.
}\label{fig:mbrelation}
\end{figure}

Figure~\ref{fig:mb2} shows the ranges of $M_\mathrm{\Ni 1}B_{14}^2$
(Eq.~\ref{mimiccond})
as a function of time for the cases of $a=2$ and 3. A factor of $2$ difference in luminosity corresponds to a 0.75-mag difference in magnitudes.
This figure shows that magnetar spin-down can mimic an energy deposition
by \Co\ decay within a factor of 2 until about 600~days
after the explosion if $M_\mathrm{\Ni 1}B_{14}^2\simeq20$.
Roughly speaking, the energy deposition from the two different energy sources
remains within a factor of 3 until about 700~days when
$10\lesssim M_\mathrm{\Ni 1}B_\mathrm{14}^2\lesssim 30$.
Figure~\ref{fig:comp} demonstrates this in the case of $M_\mathrm{\Ni 1}=10$ which
is typically found in Type~Ic SLSNe.
$B_{14}=1.4$ satisfies $M_\mathrm{\Ni 1}B_{14}^2\simeq 20$ and the
energy released by magnetar spin-down is kept within a factor of 2
of the \Co\ decay energy until about 600~days as expected.
If $B_{14}=1.7$ $(M_\mathrm{\Ni 1}B_{14}^2\simeq 30)$,
Fig.~\ref{fig:mb2}
indicates that the magnetar spin-down
and the \Co\ decay energies are kept within a
factor of 3 until about 700~days after the explosion as is seen
in Fig.~\ref{fig:comp}.
The magnetic field strength required to mimic a given amount of \Ni\ is summarized in Figure~\ref{fig:mbrelation}.

One interesting coincidence to note is that we get $B_{14}\sim 1$ for magnetars to mimic \Ni\ 
of $\sim 10~\Msun$ which are typically required for the \Ni-powered models of Type~Ic SLSNe. Because the spin-down timescale needs to be $\sim 10~\mathrm{days}$ for magnetars
to explain Type~Ic SLSN LCs, $B_{14}\sim 1$ is often found for magnetars powering Type~Ic SLSNe (\citealt{moriya2016tauris} and references therein). Therefore,
the magnetic field strengths required for Type~Ic SLSNe happen to match the required magnetic field strengths
to mimic the \Co\ decay from $\sim 10~\Msun$ of \Ni.

We have only discussed the case of the pure dipole radiation $(l=2)$ so far.
However, there are several observational indications that $l$ might be
larger than 2. For example, the braking index of the Crab pulsar is
observed to be 2.5, indicating $l=2.3$ \citep[e.g.,][]{lyne2015crab}.
A even smaller braking index around 2 is required to explain 
the peculiar SN~2005bf which is suggested to be powered by magnetar
spin-down \citep{maeda2007mag05bf}.
There are several suggested mechanisms to make the braking index
smaller, which makes $l$ larger \citep[e.g.,][]{menou2001fallback,wu2003pulsbreakindex}.

\begin{figure}
 \begin{center}
  \includegraphics[width=\columnwidth]{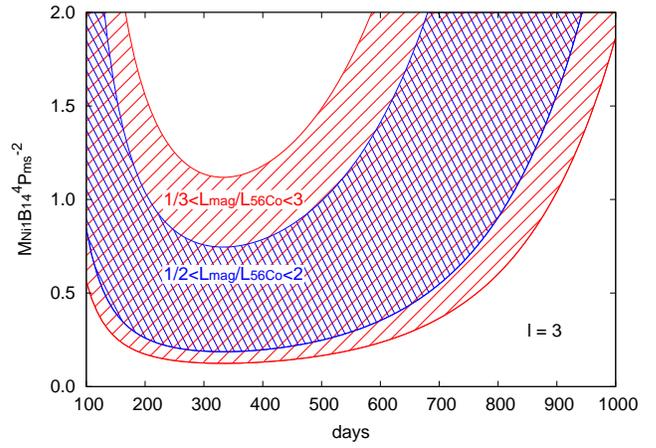} 
 \end{center}
\caption{
Ranges of $M_\mathrm{\Ni 1}B_{14}^4P_\mathrm{ms}^{-2}$
where magnetar spin-down energy with $l=3$ (a braking index of 2)
is within factors of 2 or 3 of the \Co\ decay energy.
}\label{fig:mb3}
\end{figure}

\begin{figure}
 \begin{center}
  \includegraphics[width=\columnwidth]{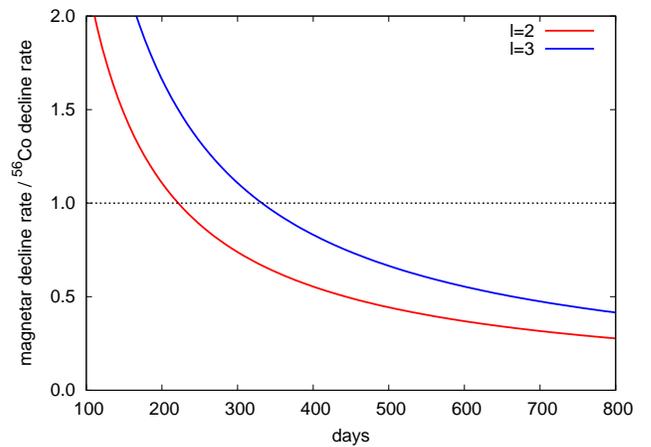} 
 \end{center}
\caption{
Ratio of energy decline rates of magnetar spin-down
and \Co\ decay. The decline rate of the \Co\ decay is constant
($0.0098~\mathrm{mag~day^{-1}}$). See also Fig.~13 of \citet{inserra2013slsntail}.
}\label{fig:grad}
\end{figure}

If magnetar spin-down has a braking index of 2 ($l=3$),
the late magnetar energy input is approximated as
\begin{equation}
L_\mathrm{mag}\rightarrow 2E_p t_p^2t^{-3}
=6.7\times 10^{63} P_\mathrm{ms}^2 B_{14}^{-4}t^{-3}~\mathrm{erg~s^{-1}},
\end{equation}
and the condition to keep the energy inputs within a factor of $a$ becomes
\begin{equation}
4.4\times 10^{20}a^{-1}e^{\frac{t}{111~\mathrm{days}}}t^{-3}
<M_\mathrm{\Ni 1}B_{14}^4P_\mathrm{ms}^{-2}<
4.4\times 10^{20}a e^{\frac{t}{111~\mathrm{days}}}t^{-3},
\end{equation}
in the cgs unit. 
Figure~\ref{fig:mb3} shows the parameter regions in which
the difference of the two energy sources is kept within
a factor of 2 or 3 in the case of $l=3$.
In the case of $l=2$, only $B_{14}$ is needed to be within a certain
range to obtain a similar energy input to the \Co\ decay.
However, in the case of $l=3$, both $B_{14}$ and $P_\mathrm{ms}$
need to be in a specific limited range satisfying 
$0.5\lesssim M_\mathrm{\Ni 1}B_{14}^4P_\mathrm{ms}^{-2}\lesssim1$
to mimic the \Co\ decay energy within a factor of 3 for a long time.
As $l$ becomes larger,
the combinations of $B_{14}$ and $P_\mathrm{ms}$
required for magnetars to mimic the \Co\ decay become more limited,
and it becomes harder for magnetars to reproduce the \Co\ decay.
Therefore, we only naturally expect the two energy sources to be similar
when magnetar spin-down occurs due to almost pure dipole radiation.
On the other hand, observed braking indices of pulsars are below 3 and they are often close to 2 or even below (e.g., \citealt{espinoza2011pulsbrakindex} and references therein). Thus, the fact that the braking indices need to be close to 3 for magnetars to mimic \Ni\ is a strong constraint on the magnetar model for SLSNe.
In the rest of this paper, we further investigate the properties of magnetars mimicking \Ni\ assuming $l=2$.
Note that the pure dipole spin-down is usually assumed in the current magnetar
spin-down models for SLSNe \citep[e.g.,][]{kasen2010iacollision,inserra2013slsntail,chatzopoulos2013anachi,metzger2014ionbreak,kasen2016magbreakout,wang2015slsnniandmag,bersten2016asassn15lh}.

Even if the energy provided by magnetar spin-down and
\Co\ decay can be kept similar for a long time,
the decline rates of the two energy sources may not remain similar for a long time.
In Fig.~\ref{fig:grad}, we show the ratio of the decline rates of the
two energy sources in the cases of
$l=2$ $(L_\mathrm{mag}\propto t^{-2})$ and
$l=3$ $(L_\mathrm{mag}\propto t^{-3})$.
The \Co\ decay rate is constant $(0.0098~\mathrm{mag~day^{-1}})$,
while the decline rates from magnetar spin-down change as a function
of time because its energy input follows a power-law.
For example, in the case of $l=2$ (Fig.~\ref{fig:comp}),
the magnetar energy input is kept within a factor of 2 of the \Co\ decay until about 600~days
after the explosion, while the LC decline rates
can differ as much as by about a factor of $\sim 2$ in
$100~\mathrm{days}\lesssim t\lesssim 600~\mathrm{days}$
(Fig.~\ref{fig:grad}).
See also \citet{inserra2013slsntail} for discussion of the decay rates.

\section{Early-phase light-curve properties}\label{sec:early}
In the previous section, we find that
the energy input from magnetar spin-down can be within a factor of 3
of the energy input from the \Co\ decay out to $\simeq 700$~days
in late phases if the condition
$10\lesssim M_\mathrm{\Ni 1}B_{14}^2\lesssim 30$ is satisfied\footnote{
We only consider the case of $l=2$ in the rest of this paper.
}.
However, even if the late phases are consistent with the \Co\ decay,
early LC properties may differ from each other.
In this section, we also look into early LC properties, namely,
rise time and peak luminosity, to
see if magnetar-powered SN LCs can mimic
\Ni-powered SN LCs even in early phases $(t\lesssim 100~\mathrm{days})$.

\subsection{Rise-time v.s. peak-luminosity relation}
\subsubsection{Magnetar-powered supernovae}\label{sec:peakmagrelation}
\citet{kasen2010mag} formulate an analytical way to estimate
the peak luminosity and rise time of SNe
powered by magnetar spin-down. We use their prescription to estimate them.
They show that the peak luminosity of magnetar-powered SNe
$(L_\mathrm{peak}^\mathrm{mag})$ for $l=2$ can be estimated as
\begin{equation}
L_\mathrm{peak}^\mathrm{mag}\simeq \frac{3E_p t_p}{2t_d^2}
\left[\ln\left(1+\frac{t_d}{t_p}\right)-\frac{t_d}{t_d+t_p}\right],
\end{equation}
where $t_d=(3\Mej\kappa/4\pi v_fc)^{0.5}$ is the effective diffusion
time determined by SN ejecta properties, i.e.,
ejecta mass \Mej, ejecta opacity $\kappa$,
characteristic final ejecta velocity
$v_f=\left[\left(E_p+\Eej\right)/2\Mej\right]^{0.5}$, and
ejecta kinetic energy \Eej.
The rise time $t_\mathrm{rise}$ for magnetar-powered SNe $(l=2)$ can be estimated as
\begin{equation}
t_\mathrm{rise}=
t_p\left(\left[
\frac{E_p}{L_\mathrm{peak}^\mathrm{mag}t_p}
\right]^{1/2}-1\right).
\end{equation}

\subsubsection{\Ni-powered supernovae}
The peak luminosity of \Ni-powered hydrogen-poor SNe ($L_\mathrm{peak}^\mathrm{\Ni}$) can be estimated
by using ``Arnett's law'' \citep{arnett1979thelaw}. For a given rise time $t_\mathrm{rise}$, the peak luminosity
roughly matches the central energy input at $t_\mathrm{rise}$. Therefore, when the central power source is \Ni\ and subsequently \Co\ decay, the peak luminosity can be estimated as
\begin{eqnarray}
L_\mathrm{peak}^\mathrm{\Ni}&=&
M_\mathrm{\Ni 1}\left[
6.5\times 10^{43}\exp\left(-\frac{t_\mathrm{rise,day}}{8.8~\mathrm{days}} \right) \right. \nonumber \\
&&\left. +1.5\times 10^{43}\exp\left(-\frac{t_\mathrm{rise,day}}{111~\mathrm{days}} \right)\right]\ \mathrm{erg~s^{-1}},
\end{eqnarray}
where $t_\mathrm{rise,day}$ is $t_\mathrm{rise}$ scaled to 1~day.

\begin{figure*}
 \begin{center}
  \includegraphics[width=\columnwidth]{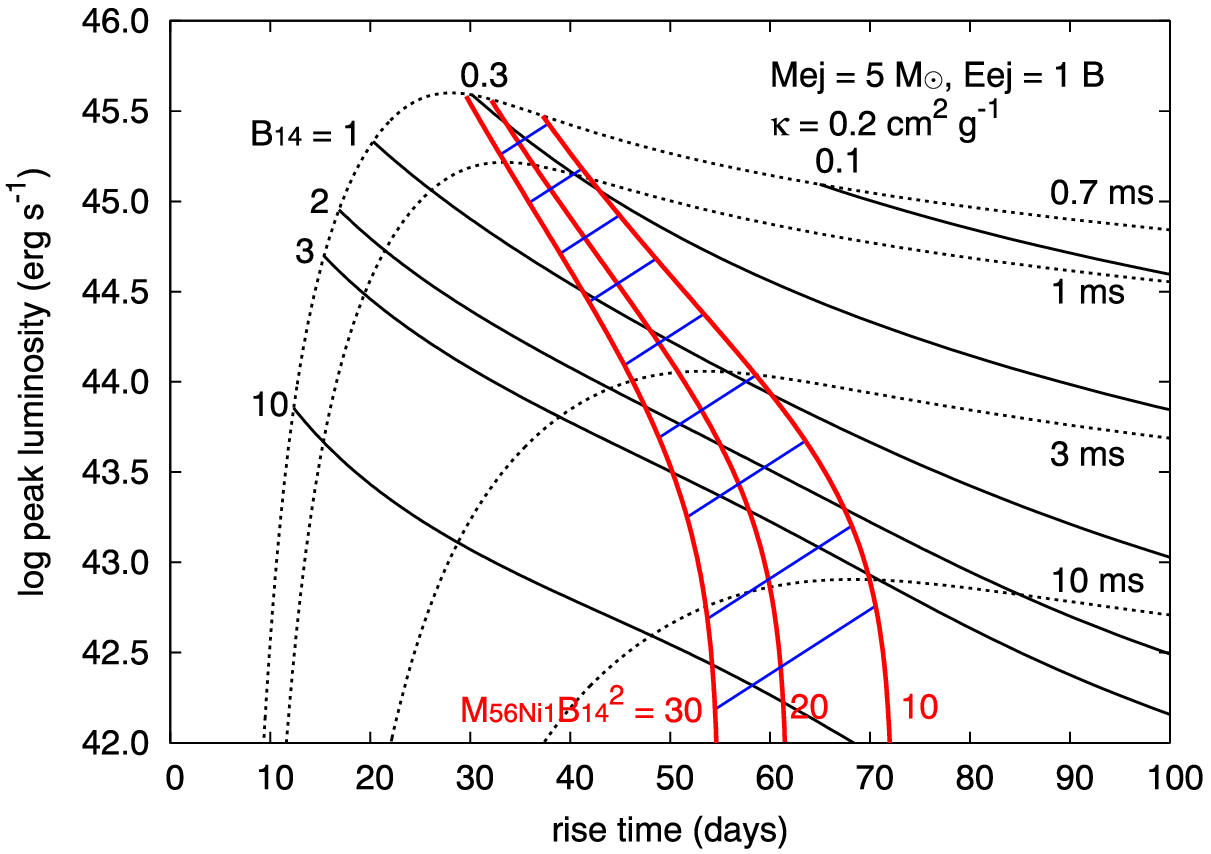} 
  \includegraphics[width=\columnwidth]{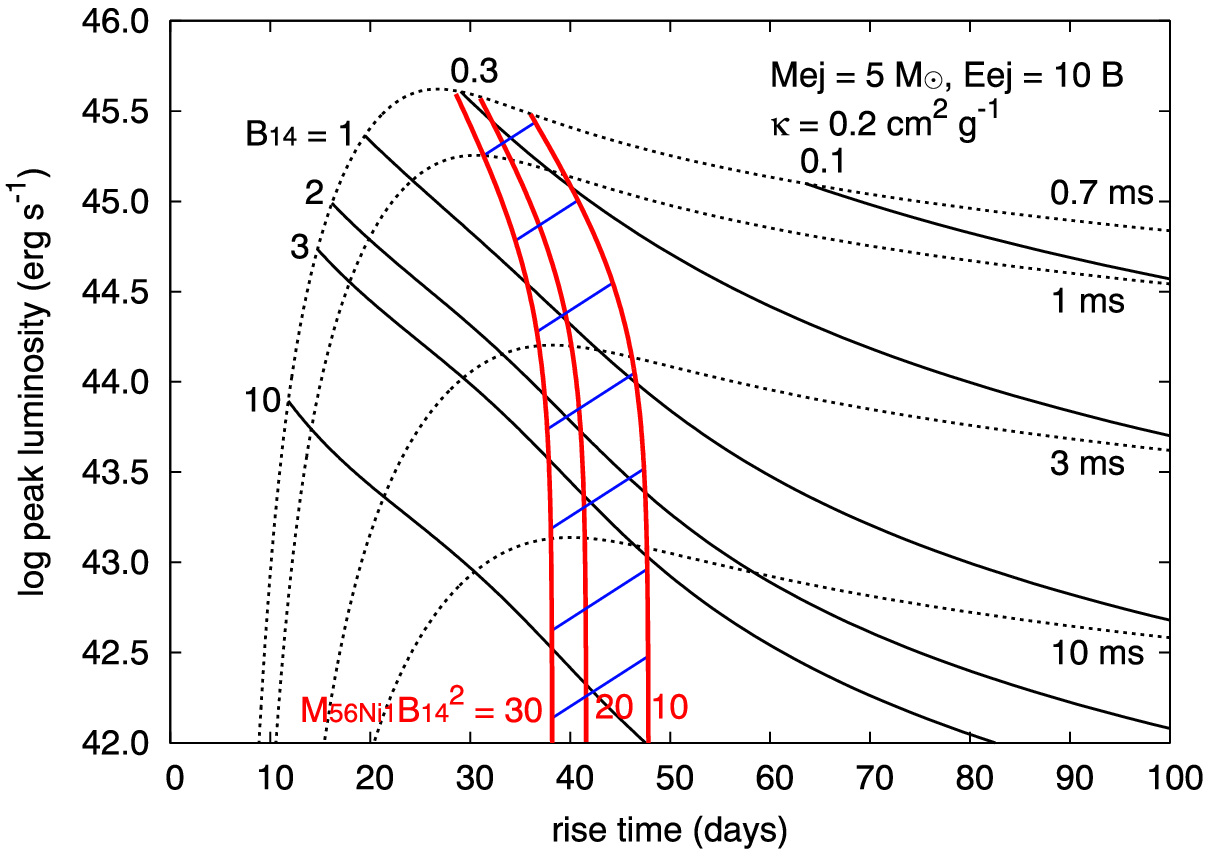} \\
  \includegraphics[width=\columnwidth]{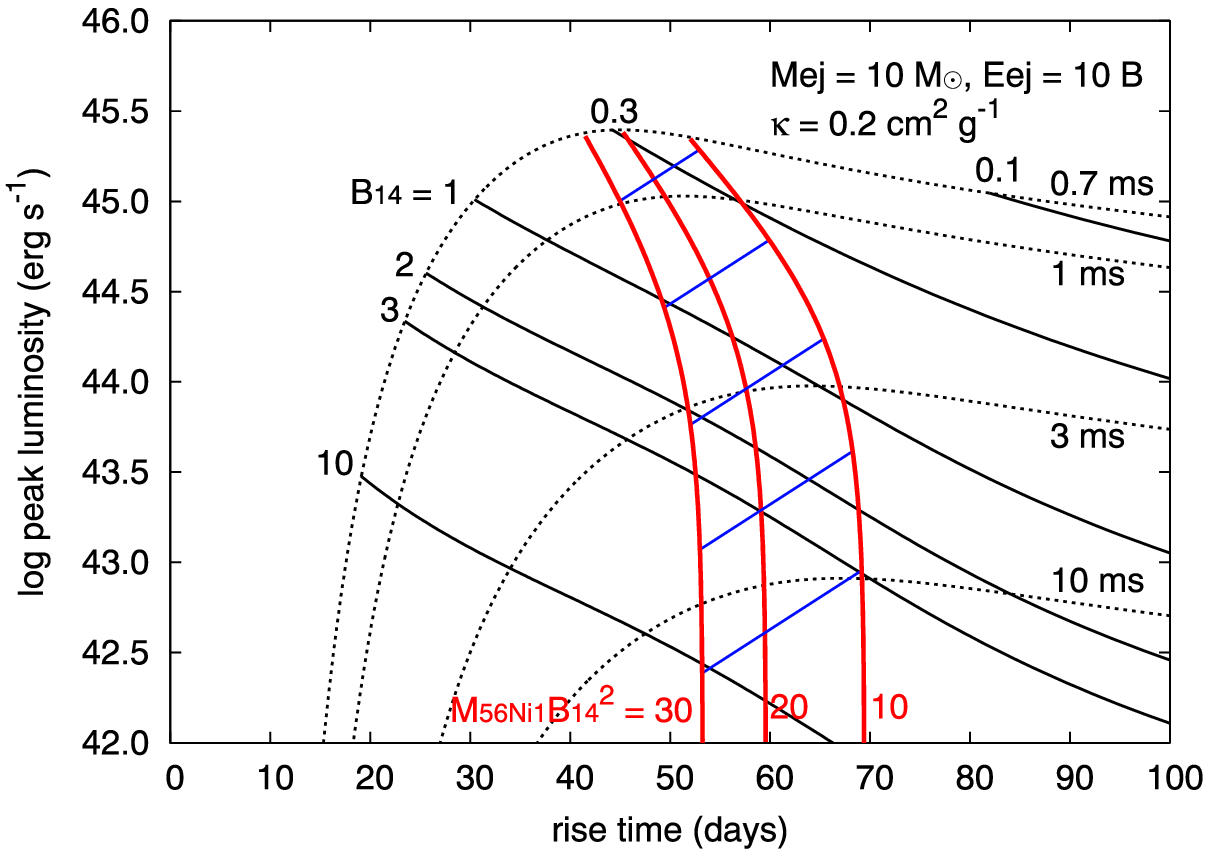} 
  \includegraphics[width=\columnwidth]{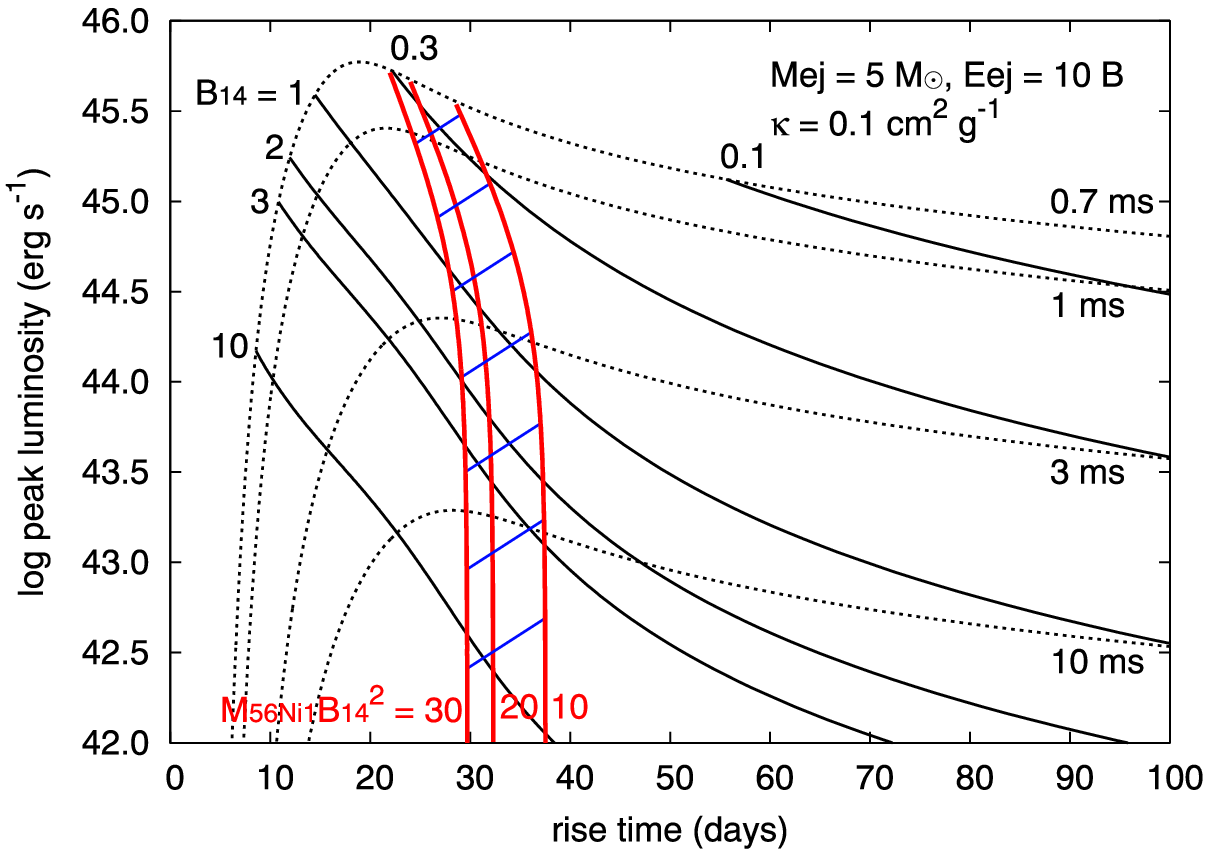} 
 \end{center}
\caption{
Rise time v.s. peak luminosity relations of SN LCs.
For the given SN ejecta properties indicated in each panels,
the required initial magnetic field strength ($B_{14}$, black solid lines) and rotational period
for magnetars (dotted lines) to obtain the rise time and the peak luminosity are plotted.
In addition, assuming a relation between rise time
and peak luminosity in \Ni-powered LCs (Eq.~\ref{eq:nimag}),
we show lines where $M_\mathrm{\Ni 1}B_{14}^2=30$, 20, and 10 are
satisfied (red solid lines).
If LC properties are in the region between the lines of $M_\mathrm{\Ni 1}B_{14}^2\simeq30$ and 10 (blue hatched),
both the magnetar-powered and the \Ni-powered models can fit
the early- and late-phase LCs at the same time.
}\label{fig:tp}
\end{figure*}

\begin{figure*}
 \begin{center}
  \includegraphics[width=1.\columnwidth]{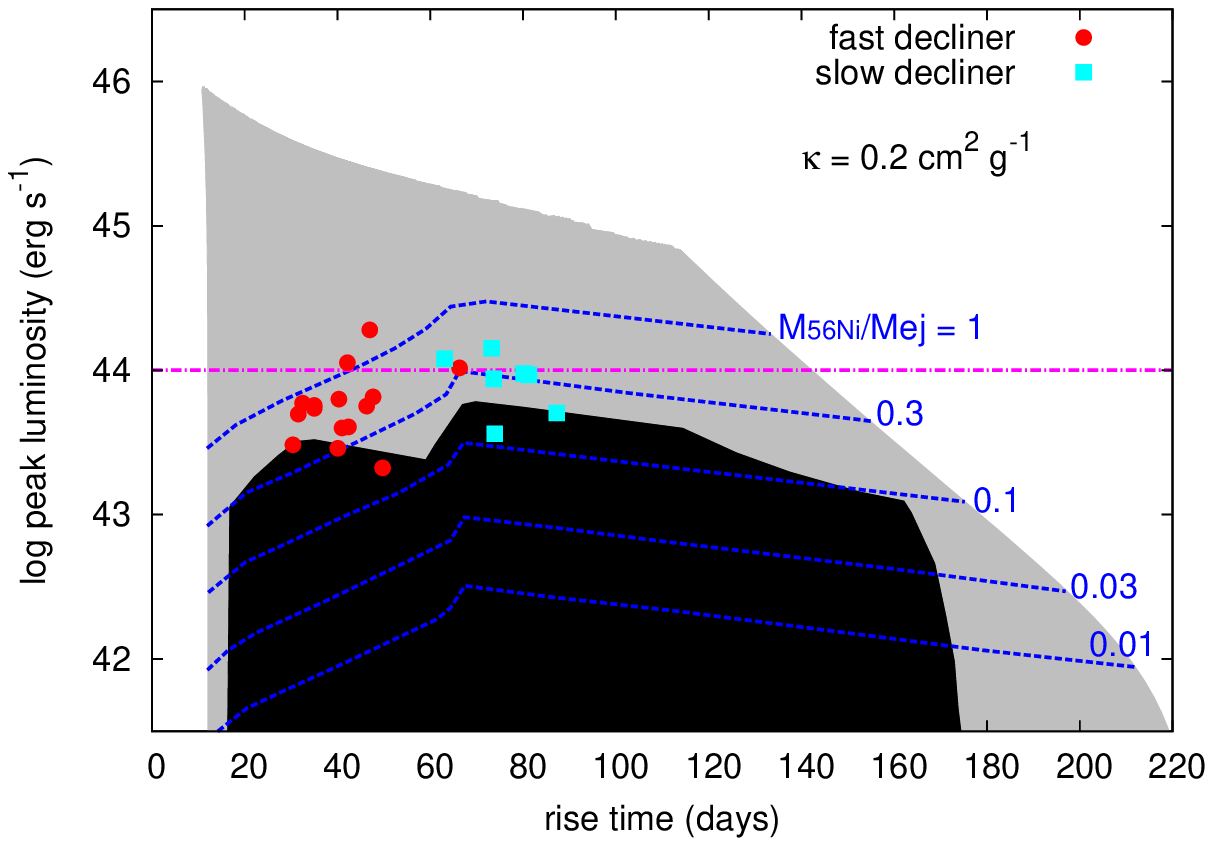} 
  \includegraphics[width=1.\columnwidth]{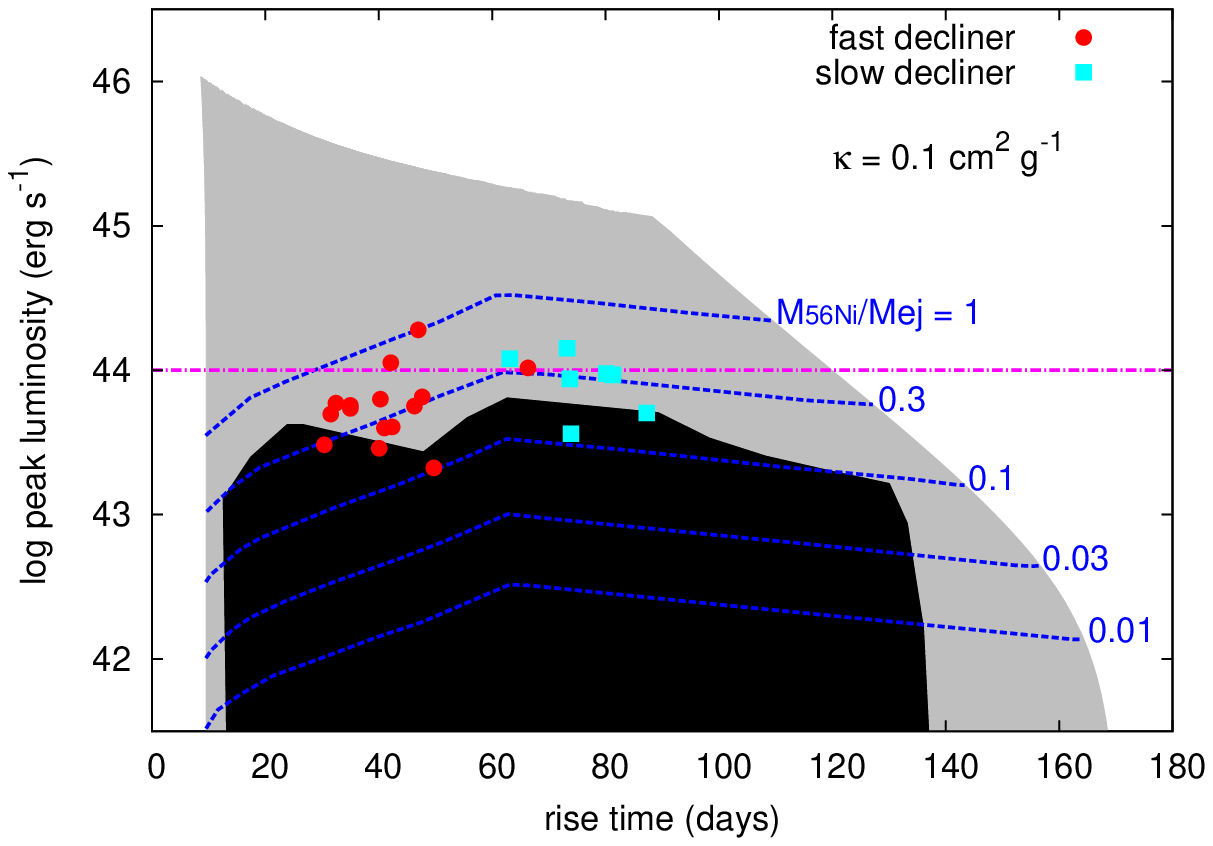} 
 \end{center}
\caption{
Regions where both magnetar-powered and \Ni-powered models
can fit early- and late-phase LCs,
for two SN ejecta opacities (left: 0.2~$\mathrm{cm^2~g^{-1}}$, and right: 0.1~$\mathrm{cm^2~g^{-1}}$).
The regions where magnetars can mimic \Ni\ from all combinations of SN ejecta
properties with $1~\Msun\leq\Mej\leq40~\Msun$ and
 $1~\mathrm{B}\leq\Eej\leq 40~\mathrm{B}$
are shown in gray.
The blue dashed lines indicates the minimum required fraction of \Ni\ mass to SN ejecta mass
to reach the peak luminosity.
The black regions are where the required \Ni\
 mass is restricted by \citet{umeda2008nomoto}, and the average SN ejecta
 velocity is between $3000~\mathrm{km~s^{-1}}$ and $30000~\mathrm{km~s^{-1}}$.
The observational properties of SLSNe summarized in \citet{nicholl2015slsndiv}
are also shown. We separate the SLSNe into two classes
(fast and slow decliners)
based on their late-phase LC decline rates as discussed in the text.
Most slow decliners have the decline rate that is consistent with the \Co\ decay within error.
The fast decliner among slow decliners is SCP06F6.
}\label{fig:magniregion}
\end{figure*}

\subsection{Mimicable rise time and peak luminosity}
Given the relations between rise time and peak luminosity in
magnetar-powered and \Ni-powered LCs, we can now investigate the parameter
range where SNe from the two power sources can have similar LC
properties in the early phases as well as in the late phases.

Rise time and peak luminosity of
magnetar-powered LCs are determined by the initial magnetic field strength
and
the initial rotational period when the SN ejecta properties are fixed
(Section \ref{sec:peakmagrelation}).
In Fig.~\ref{fig:tp}, we show the relations between 
the rise time and the peak luminosity for the magnetar spin-down model
for a given magnetic field strength and rotational period,
for several combinations of \Mej, \Eej, and $\kappa$
\citep[see also][]{kasen2010mag}.
We note that the LC modeling of Type~Ic SLSNe based on the
magnetar-powered model indicates
$\Eej\simeq 10^{51}~\mathrm{erg}\ (\equiv 1~\mathrm{B})$ 
and $\Mej\simeq 1~\Msun$ \citep[e.g.,][]{inserra2013slsntail,nicholl2014pesstoslsn,nicholl2015slsndiv}.
Their spectral modeling shows that the SN ejecta energy needs to be
at least 1~B
\citep{dessart2012magslsn}, and it is probably close to 10~B
with $\Mej \simeq 1~\Msun$ \citep{howell2013slsn}.

Given a magnetic field strength $B_{14}$, the peak
luminosity of \Ni-powered SNe satisfying $M_\mathrm{\Ni 1}B_{14}^2\equiv\alpha$
is
\begin{eqnarray}
L_\mathrm{peak}^\mathrm{\Ni}&=&
\alpha B_{14}^{-2}\left[
6.5\times 10^{43}\exp\left(-\frac{t_\mathrm{rise,day}}{8.8~\mathrm{days}} \right) \right. \nonumber \\
&&\left. +1.5\times 10^{43}\exp\left(-\frac{t_\mathrm{rise,day}}{111~\mathrm{days}} \right)\right]\ \mathrm{erg~s^{-1}}.
\label{eq:nimag}
\end{eqnarray}
Equation~(\ref{eq:nimag}) provides the $\alpha$ value required for magnetars of a given $B_{14}$ to mimic the rise time
and peak luminosity of \Ni-powered LCs during early phases.
In Fig.~\ref{fig:tp}, we plot the lines from Eq.~(\ref{eq:nimag})
for several $\alpha$.
These lines indicate the late-phase properties of magnetar-powered LCs
and they are complimentary information to the early-phase properties
in the figure which is discussed in, e.g., \citet{kasen2010mag,metzger2015magtran}.
Both the early and late LC properties in the region
between $\alpha\simeq30$ and 10 can be reproduced
by both magnetar-powered and \Ni-powered models.
Note that the \Ni\ mass required to explain the peak luminosity
by \Ni-powered models is sometimes higher than the assumed ejecta mass.
For example, at the point where the lines of $M_\mathrm{\Ni 1}B_{14}^2=20$
and $B_{14}=2$ cross, the \Ni\ mass required to account for the peak
luminosity is 5~\Msun. The required \Ni\ mass is 
as much as the SN ejecta mass in some cases.
As $B_{14}$ becomes larger, the required \Ni\ mass becomes smaller (Fig.~\ref{fig:mbrelation}).

We now vary \Mej\ and \Eej\ within a reasonable range
($1~\Msun\leq\Mej\leq40~\Msun$ and $1~\mathrm{B}\leq\Eej\leq 40~\mathrm{B}$)
and obtain the regions with $10\leq M_\mathrm{\Ni 1}B_{14}^2\leq 30$ for
all combinations of \Mej\ and \Eej\ in this range.
The gray region in
Fig.~\ref{fig:magniregion} shows where $10\leq M_\mathrm{\Ni 1}B_{14}^2\leq 30$ for $\kappa=0.2$ and $0.1~\mathrm{cm^2~g^{-1}}$.
In this region, we can find at least one combination of
$P_\mathrm{ms}$, $B_{14}$, \Mej, and \Eej\
for magnetar models to mimic \Ni-powered LCs
during both early and late phases.

To further constrain realistic parameter ranges where
magnetars can mimic \Ni, we show the minimum fraction of \Ni\ mass to SN ejecta mass
required for the \Ni-powered model. The region beyond $M_\mathrm{\Ni}/\Mej>1$, is
the ``forbidden'' region for the \Ni-powered model.
We also show the black region in Fig.~\ref{fig:magniregion}
where the following
two conditions are satisfied:
(i) the maximum \Ni\ mass for given \Mej\ and \Eej\ is restricted by those estimated by \citet{umeda2008nomoto},
and
(ii) the mean SN ejecta velocity ($\sqrt{2\Eej/\Mej}$)
is between $3000~\mathrm{km~s^{-1}}$ and $30000~\mathrm{km~s^{-1}}$.
\citet{umeda2008nomoto} estimate the maximum \Ni\ mass that can be produced by given combinations of progenitor core masses and explosion energies. Assuming that their C+O core masses roughly correspond to \Mej\ in our estimates, they show that the maximum \Ni\ mass for a given explosion energy does not change when $1\lesssim \Mej/\Msun \lesssim 20$ and then the maximum \Ni\ mass increases as \Mej\ increases. Therefore, we assume that the maximum \Ni\ mass is that of the 50~\Msun\ model (the C+O core mass of about 20~\Msun) in Fig.~7 of \citet{umeda2008nomoto} when $1\leq \Mej/\Msun \leq 20$. When $20\leq \Mej/\Msun \leq 30$, we restrict the maximum \Ni\ mass to those obtained by linearly interpolating the 50~\Msun\ and 80~\Msun\ (the C+O core mass of about 30~\Msun) models in \citet{umeda2008nomoto}. Similarly, we interpolate the maximum \Ni\ mass of the 80~\Msun\ and 100~\Msun (the C+O core mass of about 40~\Msun) models in \citet{umeda2008nomoto} when $30\leq \Mej/\Msun \leq 40$.
The condition for the mean velocity is from the estimated photospheric velocity
range of SLSNe \citep{nicholl2015slsndiv}.

To demonstrate that magnetar-powered LC properties in the mimicable region
in Fig.~\ref{fig:magniregion} can actually mimic \Ni-powered LCs,
we present a magnetar-powered LC model for SN~1998bw as an example.
SN~1998bw is a broad-line Type~Ic SN associated with GRB980425
\citep[e.g.,][]{galama1998grbsn1998bw}.
Because its early-phase spectra are rather red and
its late-phase spectra show strong Fe lines, it is likely
that SN~1998bw is a \Ni-powered SN \citep[e.g.,][]{iwamoto1998sn1998bw,mazzali2001sn1998bwnebasym}.
SN~1998bw is among the most luminous core-collapse SNe that are commonly thought to be powered by \Ni.
The rise time of SN~1998bw is $\simeq$ 15~days and the peak luminosity
is $\simeq 9\times 10^{42}~\mathrm{erg~s^{-1}}$ \citep[e.g.,][]{clocchiatti2011sn1998bw,patat2001sn1998bw},
and thus it is in the mimicable region in Fig.~\ref{fig:magniregion}.
The late-phase LC of SN~1998bw is suggested to be powered by 0.1~\Msun\
of \Ni, although the LC declines faster than that of the \Co\
decay, probably because of explosion asphericity \citep{maeda2003twocomp}.
Here, we look for a magnetar-powered LC model that has the \Ni-powered LC of SN~1998bw,
assuming $M_\mathrm{56Ni1}=0.1$.
Taking $M_\mathrm{56Ni1}B_{14}^2=15$ as the mimicking condition (Fig.~\ref{fig:mb2}),
we obtain $B_{14}=12$.

\begin{figure}
 \begin{center}
  \includegraphics[width=1.\columnwidth]{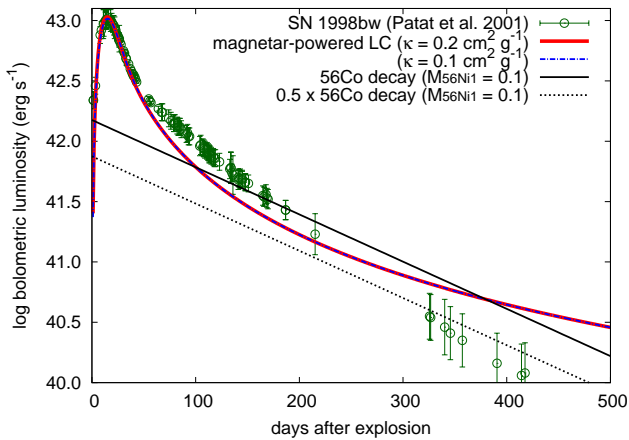} 
 \end{center}
\caption{
Magnetar-powered LC models for the \Ni-powered SN~1998bw.
The magnetar spin-down model can
successfully mimic the SN~1998bw LC for more than 500~days,
but the required ejecta mass is small ($\Mej=3-5~\Msun$).
The bolometric LC of SN 1998bw is from \citet{patat2001sn1998bw}.
}\label{fig:sn1998bw}
\end{figure}

Figure~\ref{fig:sn1998bw} shows magnetar-powered LC models for
SN~1998bw with $B_{14}=12$.
The magnetar-powered LCs are calculated in a semi-analytic way
based on \citet{arnett1982} assuming full energy trapping from
the magnetar spin-down.
This method is the same as in previous studies
of  magnetar-powered SN LCs \citep[e.g.,][]{inserra2013slsntail,chatzopoulos2013anachi}.
Our magnetar-powered LCs succeed in mimicking the \Ni-powered LC with
the given magnetic field strength in both early and late phases.
Our magnetar-powered LC models have $B_{14}=12$, $P_\mathrm{ms}=19$, and
$\Eej=30~\mathrm{B}$ in both $\kappa=0.2$ and $0.1~\mathrm{cm^2~s^{-1}}$
models.
The ejecta masses are
$\Mej=3~\Msun$ ($\kappa=0.2~\mathrm{cm^2~g^{-1}}$) and
$\Mej=5~\Msun$ ($\kappa=0.1~\mathrm{cm^2~g^{-1}}$).
The magnetar-powered LCs reproduce the early-phase LC as well as the
late-phase energy deposition from the \Co\ decay 
that powers the late-phase LC of SN~1998bw within a factor of 2
for more than 500~days.
We note that \citet{inserra2013slsntail} also show a magnetar-powered model for SN~1998bw.
They independently found a model with $B_{14}\simeq 10$,
which is close to our $B_{14}\simeq 12$  and 
satisfies the mimicking condition.

Although the magnetar model can fit the LC of SN~1998bw,
the required ejecta mass for the magnetar model
($\Mej=3-5~\Msun$ with $\Eej=30~\mathrm{B}$) is smaller than those
estimated by the LC and spectra
($\Mej\simeq 10~\Msun$ with $\Eej\simeq 30~\mathrm{B}$, e.g.,
\citealt{iwamoto1998sn1998bw,nakamura2001sn1998bw,mazzali2001sn1998bwnebasym}).
For this larger ejecta mass the magnetar model lies outside the parameter space where it can mimic the \Co\ decay.
This implies that SN~1998bw is not in the mimicable range for 
magnetars with the estimated ejecta mass and energy.
Thus, we can decline the magnetar model for SN~1998bw based on
the independent estimate for the explosion energy and the ejecta mass.

\section{Discussion}\label{sec:discussion}
\subsection{Type Ic SLSNe}
\citet{nicholl2015slsndiv} summarize the rise time and the quasi-bolometric peak luminosity
of well-observed SLSNe. They estimate the time required for SLSNe to reach the peak luminosity
($L_\mathrm{peak}$) from $L_\mathrm{peak}/e$. In Fig.~\ref{fig:magniregion},
we set the rise time as the time required to evolve from
$0.1L_\mathrm{peak}$ to $L_\mathrm{peak}$
assuming an exponential luminosity increase.
We also note that uncertain bolometric corrections are applied for some
SLSNe to estimate the peak luminosity.
We further divide SLSNe based on their decline rates.
\citet{nicholl2015slsndiv} estimate the characteristic fading time
$\tau_\mathrm{dec}$ of SLSNe,
i.e., the time required to be $L_\mathrm{peak}/e$ from
$L_\mathrm{peak}$. Their SLSNe can be separated into two groups at
$\tau_\mathrm{dec}=50~\mathrm{days}$. We call SLSNe with
$\tau_\mathrm{dec}<50~\mathrm{days}$ fast decliners and those with
$\tau_\mathrm{dec}>50~\mathrm{days}$ slow decliners.
For example, the PISN candidate SN~2007bi which shows a LC decline
consistent with the \Co\ decay soon after the peak \citep[e.g.,][]{gal-yam2009sn2007bi}
has $\tau_\mathrm{dec}\simeq8 5~\mathrm{days}$ \citep{nicholl2015slsndiv}.
Most slow decliners have a similar slow decline rate to SN~2007bi within observational errors
and therefore have the decline rate consistent with the \Co\ decay.

Looking into the observed SLSNe in Fig.~\ref{fig:magniregion},
most of them are located in the gray region where magnetars can mimic the \Ni\ decay.
Thus, even if we find a LC decline consistent with the \Co\ decay, we
cannot rule out the magnetar model based just on the decline
rate. Thus, it may not be suitable to classify slowly-decaying SLSNe
as ``SLSN-R'' based on the interpretation that the slow decay is due to
radioactive decay, as suggested by \citet{gal-yam2012slsnrev}.
Because most SLSNe are outside of the black region, core-collapse SN models are hard to explain most SLSNe with $1~\Msun\leq \Mej\leq 40~\Msun$ and $1~\mathrm{B}\leq\Eej\leq 40~\mathrm{B}$ if we adopt the maximum \Ni\ masses from core-collapse SNe in \citet{umeda2008nomoto}. The \Ni\ mass to the ejecta mass ratio typically needs to be more than 0.3 to explain SLSNe by \Ni. Relatively less luminous SLSNe near the top end of the black region in Fig.~\ref{fig:magniregion} can still be explained by core-collapse SNe \citep[cf.][]{moriya2010sn2007bi}.

The black region in Fig.~\ref{fig:magniregion} is obtained by allowing a
large range in \Mej\ and \Eej. However, once \Mej\ and \Eej\ are fixed,
the mimicable range is limited as is shown in Fig.~\ref{fig:tp}.
As we demonstrated in the previous section, the small ejecta mass
required for the magnetar-powered model disfavors this
scenario for SN~1998bw.
Thus, it is important to estimate \Mej\ and \Eej\ independently from
the power sources by using spectra to distinguish
\Ni-powered LCs from magnetar-powered LCs in both early and late phases.
\citet{nicholl2015slsndiv} have tried to estimate the ejecta masses, but
the current large uncertainty in the mass estimates prevents us
from clearly distinguishing the two sources.

As SN LCs tend to be rather symmetric when diffusion in the SN
ejecta shapes them, it is not surprising that rapidly-rising SLSNe tend
to be fast decliners and slowly-rising SLSNe tend to be slow decliners.
However, it is still interesting to note that
rapidly rising SLSNe are always fast decliners.
This is because we expect slowly-declining LCs from the magnetar model
even if the rise time is small.
However, all SLSNe with short rise times in the mimicable region
are observed as fast decliners (Fig.~\ref{fig:magniregion}).
If many slowly-declining SLSNe are actually from magnetars
mimicking the \Co\ decay, we would also expect to observe rapidly-rising SLSNe
with slow declines.
Meanwhile, we also find that magnetar-powered rapidly-rising SLSNe with slow declines
with larger \Mej\ tend to have smaller $P_\mathrm{ms}$. Therefore, the lack of
rapidly-rising slowly-declining SLSNe may otherwise indicate that SLSNe powered by 
magnetars with faster rotations tend to have smaller ejecta mass.

\subsection{Magnetar v.s. \Ni}
Even rise times of $\sim 60-90$~days which are relatively long in
the SLSN sample in Fig.~\ref{fig:magniregion} are too short to correspond to PISNe
(\citealt{kasen2011pisn,dessart2013pisn,kozyreva2014pisnlc,chatzopoulos2015rotpisn}
but see also \citealt{kozyreva2016rapidpisn}).
On the contrary, many slowly-declining SLSNe have decline rates
which are surprisingly similar to that of the \Co\ decay as discussed in Section~\ref{sec:introduction}.
We have shown in this study that
only the initial magnetic field strength needs to be within a certain range
for magnetars to mimic the \Co\ decay and this can be simply a result
of a similar initial magnetic field strength in magnetars powering SLSNe.
However, it is important to note that 
there are several mechanisms by which \Ni-powered SLSNe can have short rise times.
Strong \Ni\ mixing in PISNe can result in the short rise times \citep{kozyreva2015pisnmix},
but multi-dimensional PISN simulations do not find strong mixing in
PISNe, especially in the hydrogen-poor progenitors we are interested in \citep{joggerst2011multidpisn,chatzopoulos2013multidpisn,chen2014multidpisn}.

We have discussed one way to distinguish the two models in the previous
section: constraining \Mej\ and \Eej.
Another way to distinguish the two scenarios is
to follow the LCs for more than $\simeq 700$~days
because the magnetar spin-down energy
input will eventually becomes much larger than the \Co\ decay energy
input (Figs.~\ref{fig:mb2} and \ref{fig:comp}, see also \citealt{inserra2013slsntail}).
In addition, although the two power sources can have a similar
energy deposition for a long time, the energy decline rates from
magnetar spin-down change with time (Fig.~\ref{fig:comp}).
For example, the bolometric LC of the frequently observed SN~1987A 
clearly follows the \Co\ decay without significant deviations in 
the decline rate until about 1000~days after the explosion
\citep[e.g.,][]{seitenzahl2014sn1987a}. Magnetars cannot mimic \Ni\ for this
long, and SN~1987A is clearly not powered by magnetar spin-down.
However, it is difficult to obtain high quality
late-phase bolometric LCs for SLSNe which typically appear at 
high redshifts \citep[e.g.,][]{quimby2013slsnrate}.
In addition, as seen in the LC of SN~1998bw (Fig.~\ref{fig:sn1998bw}),
even \Ni-powered LCs may not exactly follow the \Co\ decay
because of asphericity.
Late-phase spectra can also be a way to distinguish them
\citep[e.g.,][]{mazzali2001sn1998bwnebasym,gal-yam2009sn2007bi,dessart2012magslsn,jerkstrand2016pisnnebular,jerkstrand2016slsnnebular,nicholl2016sn2015bnneb}.
Observations in X-rays and $\gamma$-rays may also distinguish the two energy sources
\citep[e.g.,][]{levan2013slsnxray,metzger2014ionbreak}.

We note that even if there is large energy input from magnetar
spin-down, it does not likely
result in a sufficiently large production of \Ni\ to
make the SNe superluminous
in the magnetar spin-down model \citep{suwa2015tominaga}.
Likely, only small amounts of \Ni\ ($\sim 0.1~\Msun$ or less)
are
synthesized during the explosion in 
magnetar-powered SLSNe \citep[cf.][]{chen2013sn2010gxhost}. 

Finally, it is impossible to explain SLSNe that have
a peak luminosity exceeding about $10^{45}~\mathrm{erg~s^{-1}}$ by \Ni\
in the parameter range in Fig.~\ref{fig:magniregion}.
The required \Ni\ mass needs to exceed the ejecta mass in this region.

\subsection{Effect of opacity}
We have compared the intrinsic energy deposition rates of the magnetar
spin-down and the \Co\ decay in this study. However, not all the deposited energy
is necessarily absorbed by SN ejecta.
In the case of the \Co\
decay, mainly $\gamma$-rays from the \Co\ decay are absorbed in
SN ejecta to power late-phase LCs. The effective $\gamma$-ray opacity
in SN ejecta is estimated to be $0.027~\mathrm{cm^2~g^{-1}}$
\citep{axelrod1980,sutherland1984wheeler}.
On the other hand, opacity for high-energy photons
created by electron-positron pairs from 
magnetar spin-down in SN ejecta is poorly investigated
\citep{kotera2013magremlc,metzger2014ionbreak}.
\citet{chen2015ptf12damhost} found that the magnetar model for 
the slowly-declining Type Ic SLSN PTF12dam cannot account for its late-phase
LC if all of the magnetar spin-down energy is deposited in the SN ejecta.
They found that the magnetar model requires $0.01$~$\mathrm{cm^2~g^{-1}}$
to match the late-phase LC (see also \citealt{wang2015slsnniandmag}).
This $\gamma$-ray opacity is similar to the $\gamma$-ray opacity for the \Co\ decay.
If the $\gamma$-ray opacity is similar in the two energy
sources, LCs from the two energy sources are expected to be similar
once the intrinsic energy inputs from the two sources are similar.
Even if the optical depth in the magnetar spin-down model is different from
the \Ni-decay model, a different $B_{14}$ that is scaled by the
difference in the opacity is still likely to make the two energy sources produce
similar LCs over long timescales.

\section{Conclusions}\label{sec:conclusions}
We have shown that magnetars have a large parameter range
where they can mimic the \Ni\ decay energy input in both early and late phases of SLSNe.
Only the initial magnetic field strength of magnetars needs to be within a certain
range ($10\lesssim M_\mathrm{\Ni 1}B_{14}^2\lesssim 30$)
for late-phase LCs powered by mangetars
to be similar to those powered by \Co\ decay out to several hundreds of days after the LC peak.
Magnetars require $B_{14}\sim 1$ to mimic \Ni\ of $\sim 10~\Msun$ which is required for SLSNe and
this magnetic field strength corresponds to those expected by SLSN LC durations.
This condition only holds if magnetar spin-down occurs by almost pure dipole
radiation and the braking index is close to 3.
As the braking index decreases,
the parameter range for magnetars to behave like \Ni\ becomes more
limited and magnetars are less likely to be able to mimic the \Ni\ decay.
Because magnetars can mimic the \Co\ decay in late phases rather
easily, it may not be appropriate to classify SLSNe as ``SLSN-R''
assuming that the slow LC decay is from radioactive decay.
With a proper combination of \Mej\ and \Eej, we can obtain 
magnetar-powered SN LCs mimicking \Ni-powered SN LCs
both in the early and late phases (Fig.~\ref{fig:tp}).

The region where magnetars can mimic \Ni\ in Fig.~\ref{fig:magniregion} indicates that
there can be slowly-declining SLSNe with short rise times.
However, all the rapidly rising SLSNe found so far also have rapid declines as shown in Fig.~\ref{fig:magniregion}. If SLSNe are powered by magnetars,
rapidly rising SLSNe with slow declines should be observed.

The range of the parameters where magnetars can mimic \Ni\
is large if there are no constraints on \Mej\ and
\Eej\ (Fig.~\ref{fig:magniregion}), but once the two SN properties are
fixed, there is only a limited range in rise time and peak luminosity
where magnetar-powered LCs can mimic \Ni-powered LCs (Fig.~\ref{fig:tp}).
Thus, it is important to obtain the SN ejecta properties independently
to distinguish the two energy sources in slowly declining SLSNe.
Bolometric LCs which extend to more than about
700~days after the explosion can also distinguish the two sources.

\acknowledgments{
We thank the anonymous referee for constructive comments that improved this work.
We thank Cosimo Inserra, Stephen Smartt, and Keiichi Maeda for discussions. We also thank Patricia Schady for comments on the manuscript. TJM is supported by the Grant-in-Aid for Research Activity Start-up of the Japan Society for the Promotion of Science (16H07413).
T.-W. Chen is supported through the Sofia Kovalevskaja Award to P. Schady from the Alexander von Humboldt Foundation of Germany.
}

\bibliographystyle{apj}
\bibliography{ms}

\end{document}